# Effects of interaction of electron-donor and –accepter molecules on the electronic structure of graphene

**Rakesh Voggu[1], Barun Das[1, 2], Chandra Sekhar Rout,[1] and C. N. R. Rao[1, 2,*]**

[1] Chemistry and Physics of Materials Unit, Jawaharlal Nehru Centre for Advanced Scientific Research, Jakkur P.O., Bangalore -560 064, India.

[2] Solid State and Structural Chemistry Unit, Indian Institute of Science, Bangalore- 560012, India.

**Abstract:**

Effects of interaction of graphene with electron-donor and –acceptor molecules have been investigated by employing Raman spectroscopy. The G-band softens progressively with the increasing concentration of tetrathiafulvalene (TTF) which is an electron-donor while the band stiffens with increasing concentration of tetracyanoethylene (TCNE) which is an electron-acceptor molecule. Both TTF and TCNE broaden the G-band. The 2D-band position is also affected by interaction with TTF and TCNE. The intensity of the 2D-band decreases markedly with the concentration of either. The ratio of intensities of the 2D- and G- bands decreases with increase in TTF and TCNE concentrations. The electrical resistivity of graphene varies in opposite directions on interaction with TTF and TCNE. All these effects occur due to molecular charge-transfer, as evidenced by the observation of charge-transfer bands in the electronic absorption spectra.

---

* For correspondence: cnrrao@jncasr.ac.in, Fax: (+91)80-22082766

The novel electronic structure and properties of graphene have attracted the attention of several workers.[1-3] The electronic structure of graphene is significantly modified by electrochemical doping of holes and electrons.[4,5] In studying such effects, Raman spectroscopy has proved to be an ideal tool.[4-6] The Raman spectrum provides the best signature of characterizing graphene, in that it is not only sensitive to the number of layers but also to dopant effects.[4-6] It has been found that the Raman G-band stiffens and sharpens on electrochemical doping with holes as well as electrons. The response of the Raman 2D-band appears to be different for holes and electrons. Equally important is that the ratio of intensities of the Raman 2D- and G- band depends on doping. We considered it important to investigate the interaction of graphene with powerful electron-donor and -acceptor molecules to explore whether such interaction will bring about significant changes in the electronic structure. We have, therefore, studied the interaction of graphene with tetrathiafulvalene (TTF) which is an excellent electron-donor and tetracyanoethylene (TCNE) which is a good electron-withdrawing molecule. For this purpose, we have employed Raman spectroscopy as well as electronic spectroscopy, the former enabling us to monitor the changes with the concentration of the donor and acceptor molecules.

Graphene was prepared by the exfoliation of graphite oxide by employing the literature procedure.[7,8] Briefly, a reaction flask containing a magnetic stir bar was charged with sulphuric acid (18 mL) and fuming nitric acid (9 mL) and cooled by immersion in an ice bath. The acid mixture was stirred and allowed to cool for 20 min, and graphite microcrystals (0.5 g) was added under vigorous stirring to avoid agglomeration. After the graphite powder was well dispersed, potassium chlorate (10 g) was added slowly over 5 min to avoid sudden increases in temperature. The reaction flask was loosely capped to allow evolution of gas from the reaction mixture and allowed to stir for 120 h at room temperature. The resulting product was suction filtered and washed thoroughly with distilled water. The product was dried under vacuum for 24 hours. The graphite oxide so obtained was exfoliated in a furnace preheated to 1050°C under argon flow for about 30s. The graphene samples were characterized using transmission electron microscopy, atomic force microscopy and powder x-ray diffraction. The number of layers in the graphene samples prepared by us was three to four. Raman spectra were recorded with a LabRAM HR high-resolution Raman spectrometer (Horiba-Jobin Yvon) using a He−Ne laser ($\lambda$ = 632.8 nm). For Raman measurements, one milligram of the graphene sample was dispersed in 3 ml of benzene containing appropriate concentrations of TTF and TCNE and sonicated. The resulting solution was filtered through an anodisc filter (Anodisc 47, Whatman) with a pore size of 0.1 μm. Electronic absorption spectra were recorded with a Perkin–Elmer UV/VIS/NIR spectrometer. In order to study the interaction with TTF and TCNE by electronic absorption spectra, the compounds were added to a suspension of graphene in acetonitrile. These suspensions were drop-coated on a quartz plate and dried. Electrical resistivity measurements were carried out by drop-coating the graphene sample on Au-gap electrodes patterned on glass substrates.

In Figure 1 we show the Raman G-bands on interaction of graphene with varying concentrations of TTF and TCNE. With increase in the concentration of TTF, there is softening of the G-band, while there is stiffening of the G-band with increasing the concentration of TCNE. The G-band broadens with increase in the concentration of either TTF or TCNE. In Figure 2(a) we show the variation in the position of the G-band maximum with the variation in concentrations of TTF and TCNE. The figure clearly shows how interaction with TTF and TCNE causes shifts in the opposite directions, the magnitude of the shift increasing with concentration. This is in contrast to the stiffening observed with electron and hole doping by electrochemical means.[4,5] Interestingly, the full-width at half maximum (FWHM) of the G-band increases on interaction with both TTF and TCNE, as can be seen from Figure 2(b). On the other hand, the G-band sharpens on electrochemical hole or electron doping.

The position of the 2D-band of graphene varies on interaction with TTF and TCNE, the latter causing an increase in the frequency. The intensity of the 2D-band decreases markedly on interaction with TTF or TCNE as shown in Figure 3(a). The ratio of intensities of the 2D and G bands, (I(2D)/I(G)), decreases markedly with increasing concentration of TTF and TCNE as shown in Figure 3(b). These results clearly reveal how the Raman spectrum of graphene is sensitive to molecular charge transfer from electron-donor and -acceptor molecules.

We have measured the I-V characteristics of graphene after interaction with different concentrations of TTF and TCNE (Figure 4). The I-V plot always remains linear, but the slope decreases with increasing concentration of TTF and increases with the increase with the concentration of TCNE. Thus, the resistance increases with increasing TTF concentration, and decreases with increasing TCNE concentration. The variation of resistance with the concentration of TTF and TCNE is shown in the inset of Figure 4(a).

We consider the changes in the electronic structure and the properties of graphene brought about by TTF and TCNE to be due to molecular charge-transfer by TTF and TCNE. We therefore, examined the UV-visible absorption spectra of graphene with the varying concentration of TTF and TCNE to look for evidence for charge transfer. TTF has a strong absorption band in the region of 305 to 316 nm, a shoulder at 361 nm and a broad band around 445 nm. The charge- transfer band of TTF with aromatics is in the 400-700 nm region.[9] We find a broad band in the 500-800 nm region in the TTF-graphene system. TCNE has a strong absorption band in the 250-270 nm region, while the charge transfer band of TCNE with aromatics is in the 550-750 nm region.[10,11] On interaction of TCNE with graphene, we observe a broad charge-transfer band between 520 and 800nm. In addition, bands possibly due to radical anions of TCNE seem to appear.[11]

In conclusion, the present results establish that molecular charge-transfer between graphene and electron-donor or -acceptor molecules causes marked changes in the electronic structure and hence in the Raman and electronic spectra as well as electrical resistivity. Eventhough we have performed our experiments with 3-4 layer graphene, the conclusions from the study are likely to be valid for single layer graphene as well.

## Figure captions:

Figure 1: Shifts in the Raman G-band of graphene on interaction with varying concentrations of TTF and TCNE

Figure 2: Variation in (a) the position of the Raman G-band and (b) in the FWHM of the G-band with the concentration of TTF and TCNE

Figure 3: Variation in the (a) Raman 2D-bands and the (b) 2D/G intensity ratio of graphene with the concentration of TTF and TCNE.

Figure 4: I-V characteristics of graphene on interaction with different concentrations of (a) TTF and (b) TCNE. Inset in (a) shows the variation in the resistance with the concentration of TTF and TCNE at a bias voltage of 0.5 V

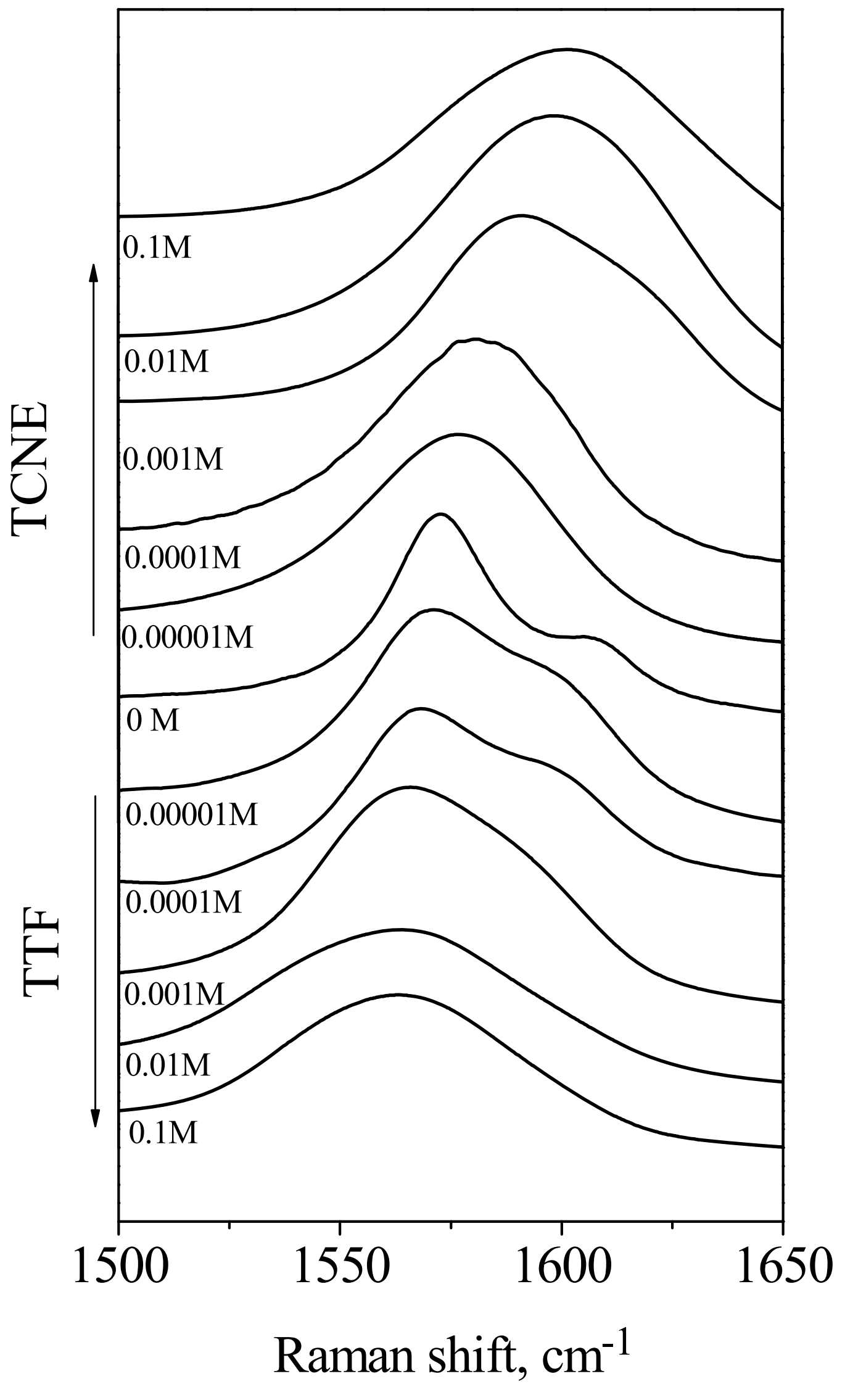

**Figure 1**

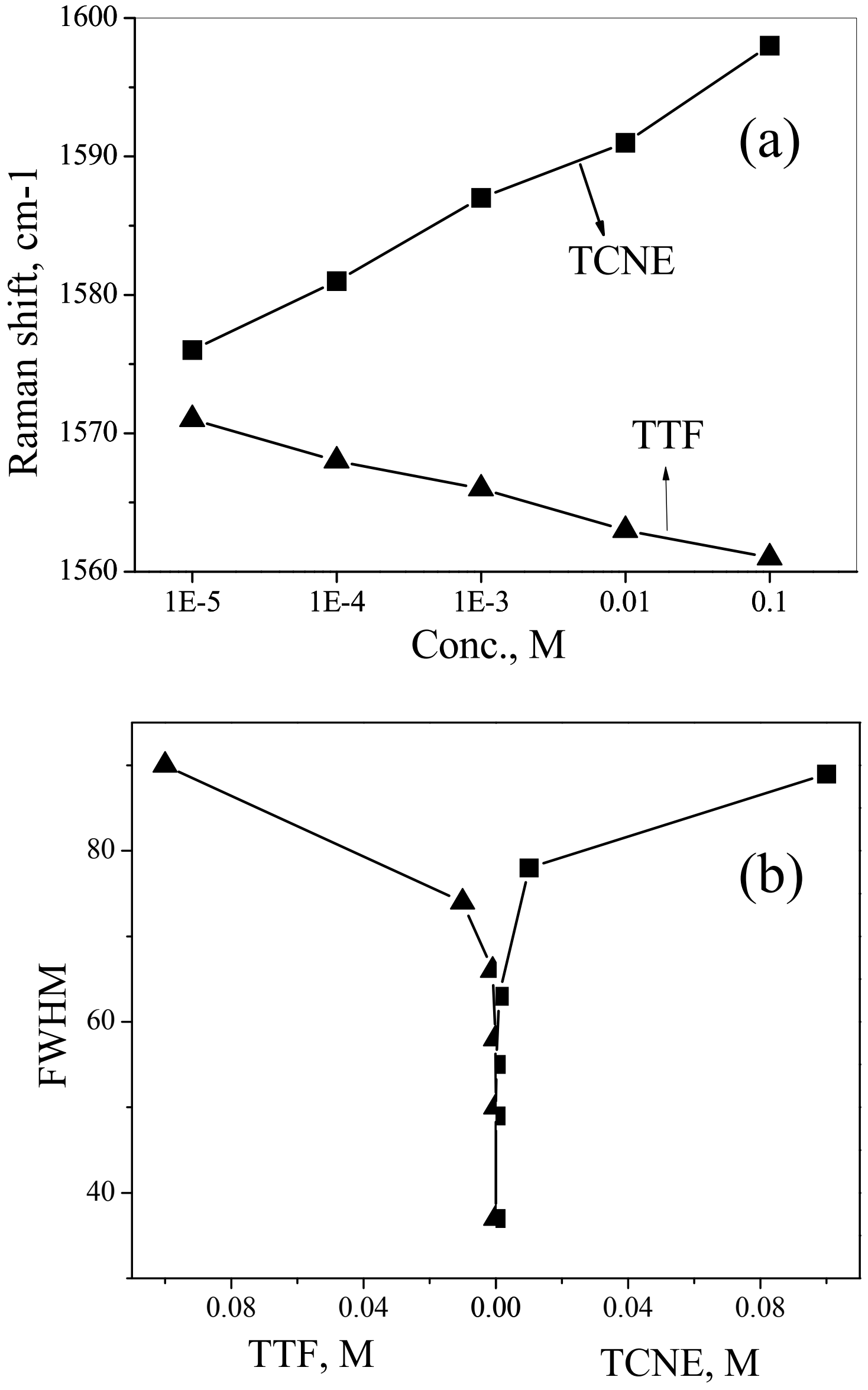

**Figure 2**

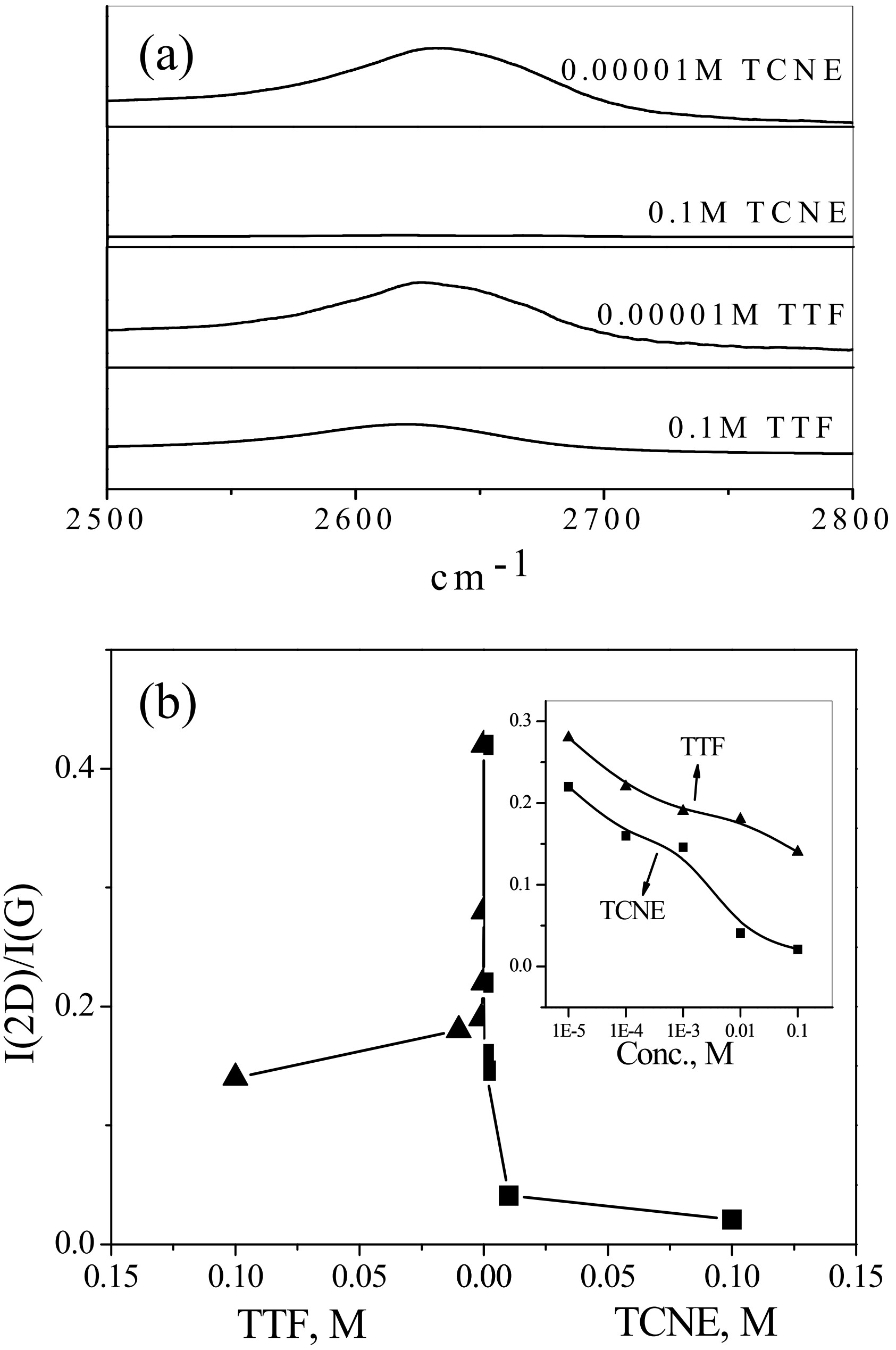

**Figure 3**

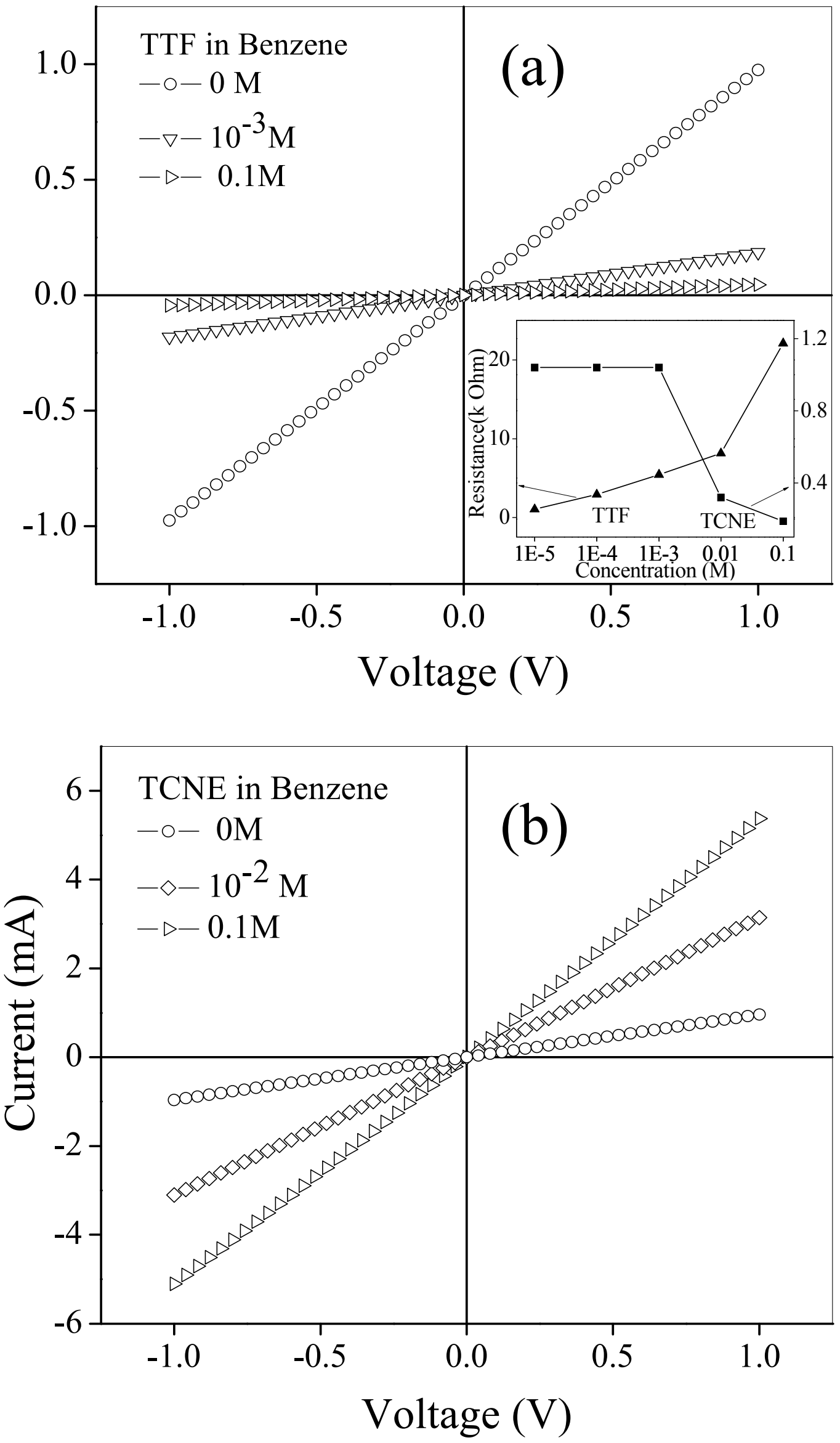

**Figure 4**